\documentclass[twocolumn,english]{revtex4-1}
\pdfoutput=1
\usepackage[T1]{fontenc}
\usepackage{color}
\usepackage{amsmath}
\usepackage{amssymb}
\usepackage{graphicx}
\usepackage{esint}
\usepackage{lineno}
\usepackage[normalem]{ulem}
\usepackage{babel}
\usepackage{units}
\usepackage{verbatim}

\makeatletter

\renewcommand{\fnum@figure}{\textbf{Figure~\thefigure}}

\makeatother

\begin{document}

\title{Interlayer exciton dynamics in van der Waals heterostructures}

\author{Simon Ovesen$^1$}
\email{simon.ovesen@chalmers.se}
\author{Samuel Brem$^1$}
\author{Christopher Linder{\" a}lv$^1$}
\author{Mikael Kuisma$^1$$^,$$^2$}
\author{Paul Erhart$^1$}
\author{Malte Selig$^3$}
\author{Ermin Malic$^1$}
\affiliation{$^1$Chalmers University of Technology, Department of Physics, 41296 Gothenburg, Sweden}
\affiliation{$^2$University of Jyv\"askyl\"a, Department of Chemistry, Nanoscience Center, 40014 Jyv\"askyl\"a, Finland}
\affiliation{$^3$Technical University Berlin, Institute of Theoretical Physics, 10623 Berlin, Germany}

\begin{abstract}
  Exciton binding energies of hundreds of $\unit{meV}$ and strong light absorption in the optical frequency range make transition metal dichalcogenides (TMDs) promising for novel optoelectronic nanodevices. In particular,  atomically thin TMDs can be stacked to heterostructures enabling the design of new materials with tailored properties. The strong Coulomb interaction gives rise to interlayer excitons, where electrons and holes are spatially separated in different layers. In this work, we reveal the microscopic processes behind the formation, thermalization and decay of these fundamentally interesting and technologically relevant interlayer excitonic states. In particular, we present for the exemplary MoSe$_2$-WSe$_2$ heterostructure the interlayer exciton binding energies and wave functions as well as their time- and energy-resolved dynamics. Finally, we predict the dominant contribution of interlayer excitons to the photoluminescence of these materials. 
\end{abstract}
 
\maketitle

\begin{figure}[b!]
    \centering
    \includegraphics[width=.60\linewidth]{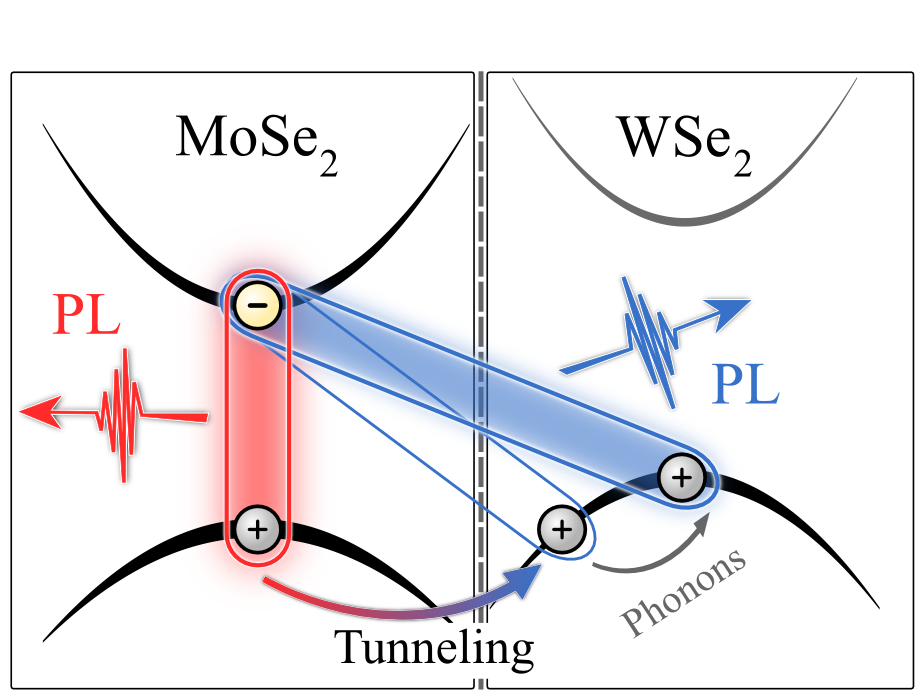}
    \caption{\textbf{Formation of interlayer excitons.} After optically exciting a coherent intralayer exciton (IaX, red oval), incoherent excitons are formed assisted by emission and absorption of phonons. Hole tunneling subsequently converts these into incoherent interlayer excitons (IeX, blue oval). After exciton thermalization through scattering with phonons, most excitons occupy the interlayer excitonic state with the vanishing center-of mass-momentum. Both IaX and IeX decay radiatively resulting in photoluminescence (PL).}
    \label{fig:HeteroConcept}
\end{figure}

A direct band gap in the optical range, efficient electron-light coupling and a remarkably strong Coulomb interaction make transition metal dichalcogenides (TMDs) highly interesting materials for both fundamental research and technological applications \cite{Geim, OptoTMD, Shan, Xuvalley, Kis, Mueller}. Tightly bound excitons, quasi-particles of Coulomb-bound electron-hole pairs, dominate the optical response of these materials \cite{LargeEff,TBEWSe2,ErminTMD, NewSemi, Crommie, Heinz, Steinhoff, Urbaszek}. They have binding energies that are one to two orders of magnitude larger than in conventional materials \cite{Jonker, HeinzBind}. 
 As a result, excitonic features are stable at room temperature and dominate the optical response and non-equilibrium dynamics in TMDs. Besides regular bright excitons, TMDs also exhibit a variety of optically forbidden dark excitons \cite{Potemski, Potemski2, HeinzDark, HeinzDark2, Potemski3, MalteLW, MalteDE, MajaNC}, which cannot be addressed optically due to the required momentum transfer or spin flip.
The fascinating exciton physics becomes even richer, when considering that atomically thin materials can be vertically stacked to form Van der Waals (VdW) heterostructures \cite{Geim}. 
In these systems the strong Coulomb interaction gives rise to interlayer excitons, where the involved electrons and holes are located in different layers (Fig. \ref{fig:HeteroConcept}). After optical excitation of a regular intralayer exciton (IaX), the hole can tunnel to the other layer forming an interlayer exciton (IeX). 
 Due to an offset in the alignment of the monolayer band structures (type II heterostructures), these interlayer excitons lie energetically below the excitons confined within one layer \cite{Thegysen}  (Fig. \ref{fig:HeteroConcept}). 
Depending on  spin and momentum of the states involved, interlayer excitons can be either bright or dark.
 
VdW heterostructures present an emerging field of research, as evidenced by an increasing number of studies, in particular demonstrating the appearance of interlayer excitons in photoluminescence (PL) spectra \cite{SuperCond, Interex, Cao, Geim2, Hogele, KornZeeman, Low, Macdonald, Spataru, Rana}. In the low-temperature regime, a pronounced additional resonance is observed at an energy below the intralayer excitons of the single layers \cite{InterKorn, KornZeeman, Interex, RiveraPol, Miller}. The PL intensity of this low-energy peak is very pronounced compared to the intralayer exciton in the weak excitation regime. This behavior can be traced back to  formation of interlayer excitons that due to their spectral position are highly occupied. Furthermore, in time-resolved PL measurements a spectrally narrow resonance was observed exhibiting lifetimes of tens to hundreds of nanoseconds \cite{InterKorn, KornZeeman, Interex, RiveraPol, Miller}. Theoretical studies of  VdW heterostructures have  so far been restricted to static observables, such as screening \cite{Gies}, excitonic binding energies \cite{Thegysen}, and lattice mismatch effects \cite{Macdonald}.  Microscopic insights into the ultrafast dynamics of interlayer excitons have remained literally in the dark yet. 

In this work, we present a microscopic view on the exciton dynamics in vdW heterostructures, in particular revealing the time- and energy-resolved processes behind the formation, thermalization and decay of interlayer excitons. We predict the  binding energy of $\unit[173]{meV}$ for the energetically lowest interlayer excitonic state in the exemplary MoSe$_2$-WSe$_2$ heterostructure. 
 Moreover, we shed light on the exciton dynamics demonstrating that interlayer excitons are formed via hole tunneling on a sub-picosecond timescale, followed by a much slower radiative interlayer recombination. Finally, we provide a microscopic explanation for the recently performed time-resolved photoluminescence experiments demonstrating the dominant contribution of interlayer excitons to the photoluminescence of the MoSe$_2$-WSe$_2$ heterostructure. 

To provide microscopic access to the  dynamics of the coupled excitons, phonons and photons, we apply the density matrix formalism \cite{KKQE, Thranhardt2000, KuhnUltra, KochBook} and derive the luminescence Bloch equations explicitly including excitonic effects \cite{ErminTMD, Thranhardt2000}.
The emitted luminescence intensity $I_\omega(t)$ is obtained from the temporal change of the photon density and reads \cite{Thranhardt2000, KKQE2}.
$$
  I_{\omega}(t) \propto \omega \sum_{l_h l_e} | M_{\boldsymbol{0}}^{l_h l_e} |^2 \left(|P^{l_h l_e}_{\boldsymbol{0}}(t)|^2 + N^{l_h l_e}_{\boldsymbol{0}} (t) \right) \mathcal{L}_\gamma(\Delta E^{l_h l_e}_{\omega})
$$
Here,  $M_{\boldsymbol{0}}^{l_h l_e}$ is the optical matrix element for excitons  that are composed of electrons in layer $l_e$ and holes in layer $l_h$. The index $\boldsymbol{0}$ indicates that only excitons with vanishing center-of-mass momentum $\boldsymbol{Q}=\boldsymbol{0}$ contribute to PL. While the optical matrix element can be obtained analytically using a tight-binding approach for TMD monolayers and adjusting the coupling strength to experimentally measured absorption \cite{ErminTMD, MalteLW}, first-principle calculations have been performed to determine the coupling for interlayer excitons. The  Lorentzian $\mathcal{L}_\gamma$ accounts for  energy conservation, i.e.  an exciton decays into a photon with the same energy. We  calculate on a microscopic level the components of the dephasing rate $\gamma$ stemming from exciton-phonon and exciton-photon interactions \cite{MalteLW, Li}. At lower temperatures disorder-induced dephasing becomes important and has been accounted for phenomenologically by using the FMHW values measured in Ref. \cite{InterKorn}.

The PL strength is determined by the excitonic polarization  $P_{\boldsymbol{Q}}^{l_h l_e}(t)$ reflecting the optically driven coherence (often referred to as coherent excitons \cite{KKQE, Thranhardt2000}) and by the exciton occupation $N^{l_h l_e}_{\boldsymbol{Q}}$ describing the formation and thermalization of  incoherent excitons induced by a non-radiative decay of $P_{\boldsymbol{Q}}^{l_h l_e}(t)$.
The excitonic polarization is defined as 
$P_{\boldsymbol{Q}}^{l_h l_e}(t) = \sum_{\boldsymbol{q}} \varphi_{\boldsymbol{q}}^{* h e} \langle v_{\boldsymbol{q} + \beta^{l_h l_e} \boldsymbol{Q}}^{\dagger h} c_{\boldsymbol{q} - \alpha^{l_h l_e} \boldsymbol{Q}}^{e} \rangle$, where $\varphi_{\boldsymbol{q}}$ are excitonic wavefunctions in momentum space and $v^{(\dagger)}, c^{(\dagger)}$ are operators for annihilation (creation) of valence- and conduction band electrons, respectively. 
The incoherent exciton occupation can be expressed as \begin{footnotesize} $ N_{\boldsymbol{Q}}^{l_h l_e}(t)=\sum_{\boldsymbol{q,q'}} \varphi^{* h e}_{\boldsymbol{q'}} \varphi^{l_h l_e}_{\boldsymbol{q}} \delta \langle c^{\dagger e}_{\boldsymbol{q}-\alpha^{l_h l_e} \boldsymbol{Q}} v^{h}_{\boldsymbol{q}+\beta^{l_h l_e} \boldsymbol{Q}} v^{\dagger h}_{\boldsymbol{q}'+\beta^{l_h l_e} \boldsymbol{Q}} c^{e}_{\boldsymbol{q'}-\alpha^{l_h l_e} \boldsymbol{Q}} \rangle(t)$ \end{footnotesize} corresponding to electron-hole pair correlations \cite{Thranhardt2000}. Here, we have introduced relative $\boldsymbol{q}$ and center-of-mass momenta $\boldsymbol{Q}$ with the coefficients $\alpha = m_\text{e}/(m_\text{h}+m_\text{e})$ and $\beta = m_\text{h}/(m_\text{h}+m_\text{e})$ describing the relative electron and hole masses.

The excitonic eigen energies and wave functions are obtained by solving the Wannier equation \cite{ErminTMD, KKQE, KochBook}
\begin{equation}
   \frac{\hbar^2\boldsymbol{q}^2}{2\mu^{l_h l_e}} \varphi_{\boldsymbol{q}}^{l_h l_e} - \sum_{\boldsymbol{k}}V^{l_h l_e}_{\boldsymbol{k} - \boldsymbol{q}} \varphi_{\boldsymbol{k}}^{l_h l_e} = E^{l_h l_e}_{b} \varphi_{\boldsymbol{q}}^{l_h l_e},
\end{equation}
where $\mu$ is the reduced mass, $E_{b}$ the excitonic binding energy, and $V^{l_h l_e}_{\boldsymbol{k}}$ the Coulomb matrix element for the electron-hole interaction. To obtain the interaction potential between charge carriers in a heterostructure, we generalize the widely used Keldysh potential for the monolayer case \cite{ErminTMD, Keldysh}. We solve the Poisson equation for two aligned homogeneous slabs. This gives rise to an effective 2D Coulomb potential $V^{l_h l_e}_{\boldsymbol{k}} = \frac{e_0^2}{k \varepsilon_0 \varepsilon^{l_h l_e}(\boldsymbol{k})}$ with a dielectric function $\varepsilon^{l_h l_e}(\boldsymbol{k})$, depending on the momentum transfer $\boldsymbol{k}$ and the overall composition of the heterostructure, cf. supplementary material for more details. 

Equations of motion for the exciton polarization and the exciton occupations are derived taking into account all relevant interaction mechanisms in the low density regime. Specifically, we include the coupling to the exciting laser pulse  as well as the carrier-photon interaction giving rise to a spontaneous radiative decay of excitons. Moreover, excitons are coupled to optical and acoustic phonon modes, allowing for a relaxation of the excited hot exciton distribution into a thermal equilibrium. The carrier-phonon matrix elements are taken from DFT calculations \cite{Phon} and are treated in analogy to Refs. \cite{MalteLW, MalteDE,SamuelCascade}. The interaction mechanism that distinguishes the exciton dynamics in a VdW heterostructure from the dynamics in a bare TMD monolayer is the tunneling of carriers between layers. This process is included via the tunneling Hamilton operator $H_T = \sum_{a, b} T^{a b} a^{\dagger}_b a_b$, where $a,b$ are compound indices containing layer, band and momentum of the electron. The coupling element $T$ is given by the overlap integral $\left< \Psi_a | V_T | \Psi_b \right>$ of Bloch waves $\Psi$ with the interlayer potential $V_T$. The latter can be separated 
into an out-of-plane component $V_z$ and an in-plane disorder potential $V_{\boldsymbol{\rho}}$ \cite{Tunneling}. 
The first is given by a step function, which is only non-zero within the region between the two layers and its value  was fixed to $\unit[5]{eV}$ corresponding to the ionization energy of TMD monolayers \cite{EIon, EIon2}. The tunneling matrix element can then be expressed as $T^{a b} = V_{\boldsymbol{\rho}} (|\boldsymbol{k}_b - \boldsymbol{k}_a|) V_z \left< u^a | u^b \right>_{\text{uc}}$, where $u$ are the lattice-periodic parts of the Bloch waves, which are integrated over one unit cell (uc). This integral was obtained from density functional theory calculations, yielding an overlap of approximately $1 \cdot 10^{-2}$. 
The calculations were carried out using the \textsc{gpaw} package \cite{EnkRosMor10}.
The wave function was expanded on a grid and exchange-correlation effects were represented using the PBE exchange-correlation functional \cite{PerBurErn96}.

Finally, the Fourier transform of the in-plane component of the disorder potential reads \cite{Tunneling} $V_{\boldsymbol{\rho}} (|\boldsymbol{k}_b - \boldsymbol{k}_a|)=\sqrt{\pi} L_C / \left(1 + \frac{|\boldsymbol{k}_b -\boldsymbol{k}_a|^2 L_C^2}{2}\right)^\frac{3}{4}$ with  $L_C$ as the correlation length. It has been set to $\unit[1]{nm}$ in accordance with the excitonic Bohr radius \cite{GunnarDisorderSource}, an approximation that applies for short-range disorder \cite{StatPhys, QWire}. 

Applying the Heisenberg equation of motion, we obtain the luminescence Bloch equations for VdW heterostructures
\begin{align}
\nonumber
    \dot{P}_\alpha&= \frac{1}{i\hbar} E_\alpha P_\alpha+ i \Omega_\alpha - \bigg( \gamma^\alpha_{r} - \frac{1}{2} \sum_\beta \Gamma_{P}^{\alpha \beta} \bigg) P_\alpha, \\
  \nonumber
    \dot{N}_\alpha &= \sum_{\beta} \left(\Gamma_{P}^{\beta \alpha} \left|P_\beta\right|^2+ \Gamma_T^{\alpha \beta} \Delta N_{\alpha\beta}\right) - 2 \gamma^\alpha_{r}  N_{\alpha, \boldsymbol{0}} + N_\alpha^{\text{s}} ,
\end{align}
where the excitonic compound indices $\alpha, \beta$ contain center of mass momentum and electron/hole layers. The dynamics of the exciton polarization $P_\alpha(t)$ is determined by the Rabi frequency $\Omega_\alpha$ containing the driving optical pump pulse and the decay processes stemming from radiative damping ($\gamma^\alpha_r$) and electron-phonon interaction ($\Gamma^{\alpha\beta}_P$). The dynamics of the incoherent exciton occupation $N_\alpha(t)$ is determined by formation processes driven by phonon-assisted decay of the excitonic polarization, the radiative decay, and exciton-phonon scattering ($N_\alpha^{\text{scat}}$) driving the excited system towards an equilibrium Boltzmann distribution \cite{MalteDE,SamuelCascade}. Note that that radiative decay scales with $N_{\alpha, \boldsymbol{0}}=N_{\alpha}\delta_{\boldsymbol{Q,0}}$, where only states within the light cone with a nearly zero center-of-mass momentum $\boldsymbol{Q}=\boldsymbol{0}$ can contribute.  Finally, $\Gamma^{\alpha\beta}_T$ describes resonant tunneling between different layers. It depends on the occupation difference ($\Delta N_{\alpha\beta}=N_\beta - N_\alpha$) in the involved excitonic states  and causes the formation of interlayer excitons.  Details about the applied approach can be found in the supplementary material. 

\begin{figure}[t!]
\includegraphics[width=\linewidth]{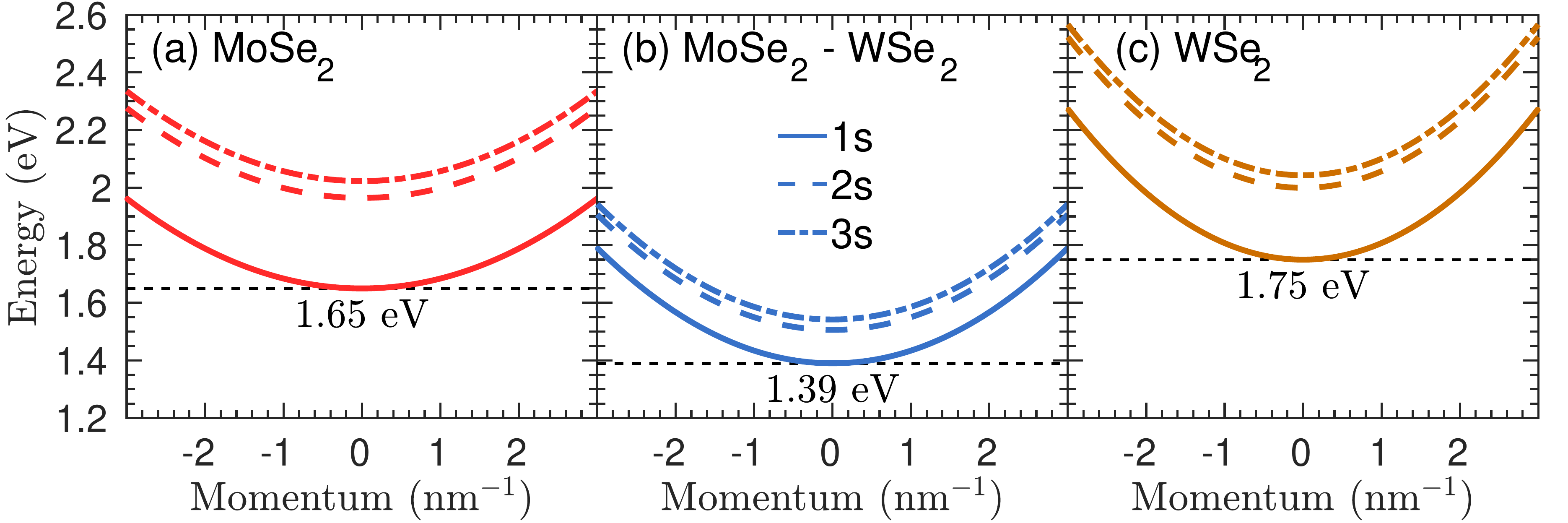}
     \caption{\textbf{Exciton band structure.} Energetically lowest excitonic states (1s, 2s, 3s) for  MoSe$_2$ and WSe$_2$ intralayer excitons and the corresponding MoSe$_2$-WSe$_2$ interlayer exciton, respectively. The corresponding excitonic wave functions are shown in in the supplementary material.}
    \label{fig:Disp}
\end{figure}

The derived Bloch equations provide microscopic access to time-, momentum- and energy-resolved formation, relaxation, and decay dynamics of intra- and interlayer excitons. In this work, we investigate the exemplary MoSe$_2$-WSe$_2$ heterostructure on a typical SiO$_2$ substrate. To bring the system into a non-equilibrium, we apply a laser pulse at the energy resonant to the 1s exciton of the MoSe$_2$ layer. Solving first the  Wannier equation, we obtain the excitonic band structure and the corresponding excitonic wave functions, which are shown for the three energetically lowest excitonic states (1s, 2s, 3s) for all intra- and interlayer excitons in Fig. \ref{fig:Disp}. Here, the electronic band alignment has been extracted from PL measurements \cite{Interex, InterKorn}. Furthermore, we explicitly account for the changed screening, when the TMD monolayers are integrated within a heterostructure, cf. the supplementary material. We find that the interlayer exciton is the energetically lowest state at $\unit[1.39]{eV}$, while the intralayer excitons lie at $\unit[1.65]{eV}$ (MoSe$_2$) and $\unit[1.75]{eV}$ (WSe$_2$), cf. Fig. \ref{fig:Disp}. The resulting excitonic binding energies are displayed in Table \ref{tab:EBind}. As expected, the binding energy for interlayer excitons is significantly reduced ($\unit[173]{meV}$ for 1s) compared to the value for intralayer excitons ($\unit[413]{meV}$ for MoSe$_2$ and $\unit[317]{eV}$ for WSe$_2$), however it is still much larger than the thermal energy. Thus, interlayer excitons are expected to be stable at room temperature and significantly contribute to the PL. 
Our calculations also reveal that the binding energy of intralayer excitons is reduced by some tens of meV when the monolayers are stacked into a heterostructure (cf. Table \ref{tab:EBind}). This is due to an increased screening of the Coulomb potential.
 \begin{figure}[t!]
\includegraphics[width=.85\linewidth]{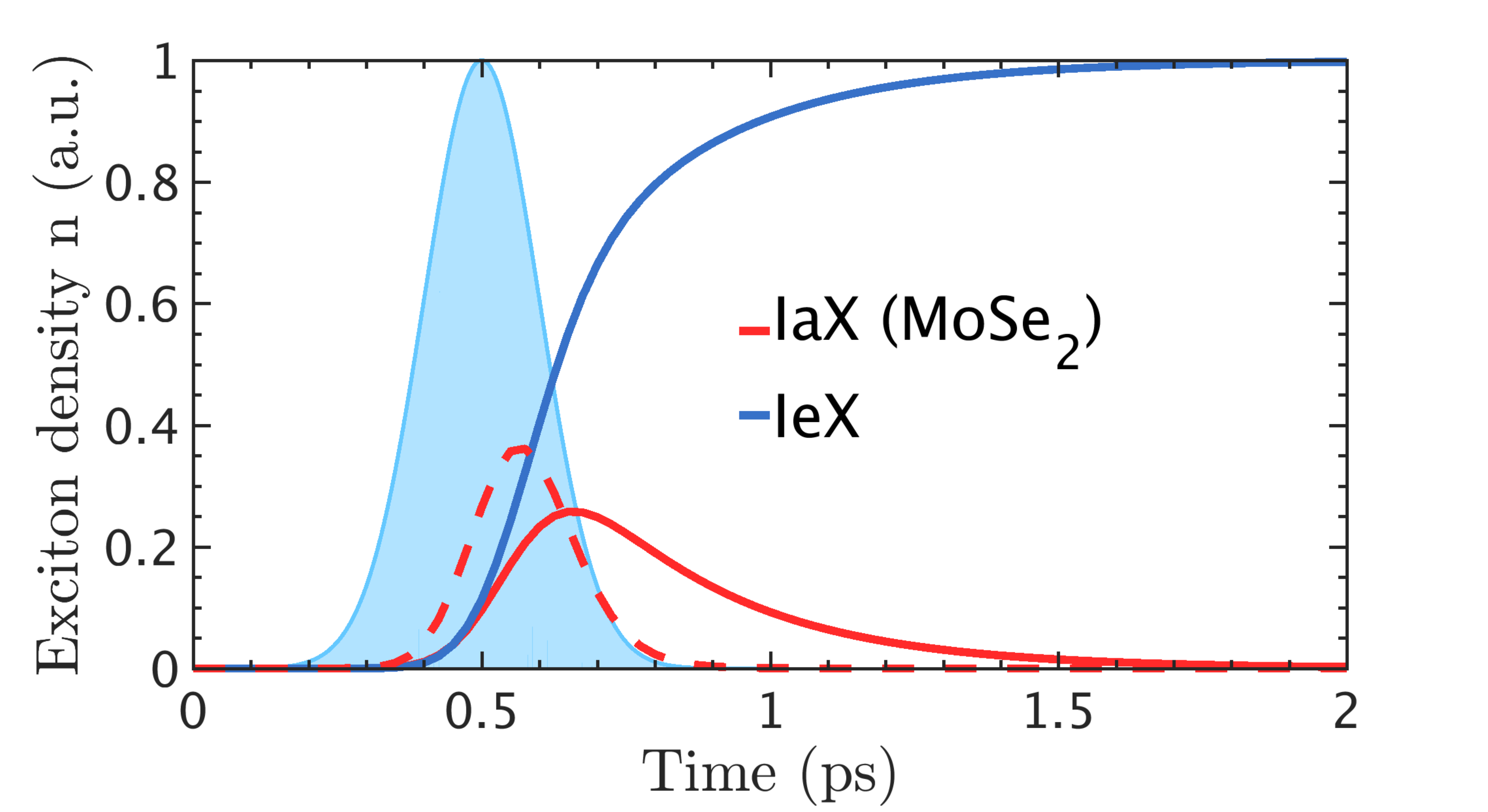}
    \caption{\textbf{Momentum-integrated exciton dynamics.} Evolution of the MoSe$_2$ intralayer (IaX) and the interlayer (IeX) 1s exciton density  at $\unit[77]{K}$ after resonant excitation of the MoSe$_2$ layer. The optical pulse is denoted by the blue-shaded Gaussian. The dashed line represents  coherent excitons in the MoSe$_2$ layer. The interlayer exciton formation occurs on a sub-ps timescale. The radiative lifetime determining the decay of exciton densities is in the order of microseconds.}
    \label{fig:NtP} 
\end{figure}

\begin{table}[b!]
    \begin{small}
    \begin{tabular}{l|c|c|c||c|c}
        \multicolumn{6}{c}{\textbf{Excitonic binding energies (meV)}} \\
        \hline
         & \multicolumn{3}{c||}{Heterostructure} & \multicolumn{2}{c}{Monolayer} \\
        \hline
         & MoSe$_2$ & WSe$_2$ & MoSe$_2$-WSe$_2$ & MoSe$_2$ & WSe$_2$ \\
        \hline
        1s & 413 & 317 & 173  & 434 & 343 \\
        2s & 111 & 77 & 69 & 127 & 86 \\
        3s & 53 & 34 & 35 & 58 & 37
    \end{tabular}
    \end{small}
     \caption{\textbf{Excitonic binding energies.} Solving the Wannier equation we have full access to all excitonic states. The  binding energy of the 1s interlayer exciton is roughly half of the intralayer excitons. For higher states the difference becomes smaller. The binding energy of intralayer excitons is reduced due to the increased screening within the heterostructure.}
     \label{tab:EBind}
\end{table}

\begin{figure}[!t]
\includegraphics[width=0.85\linewidth]{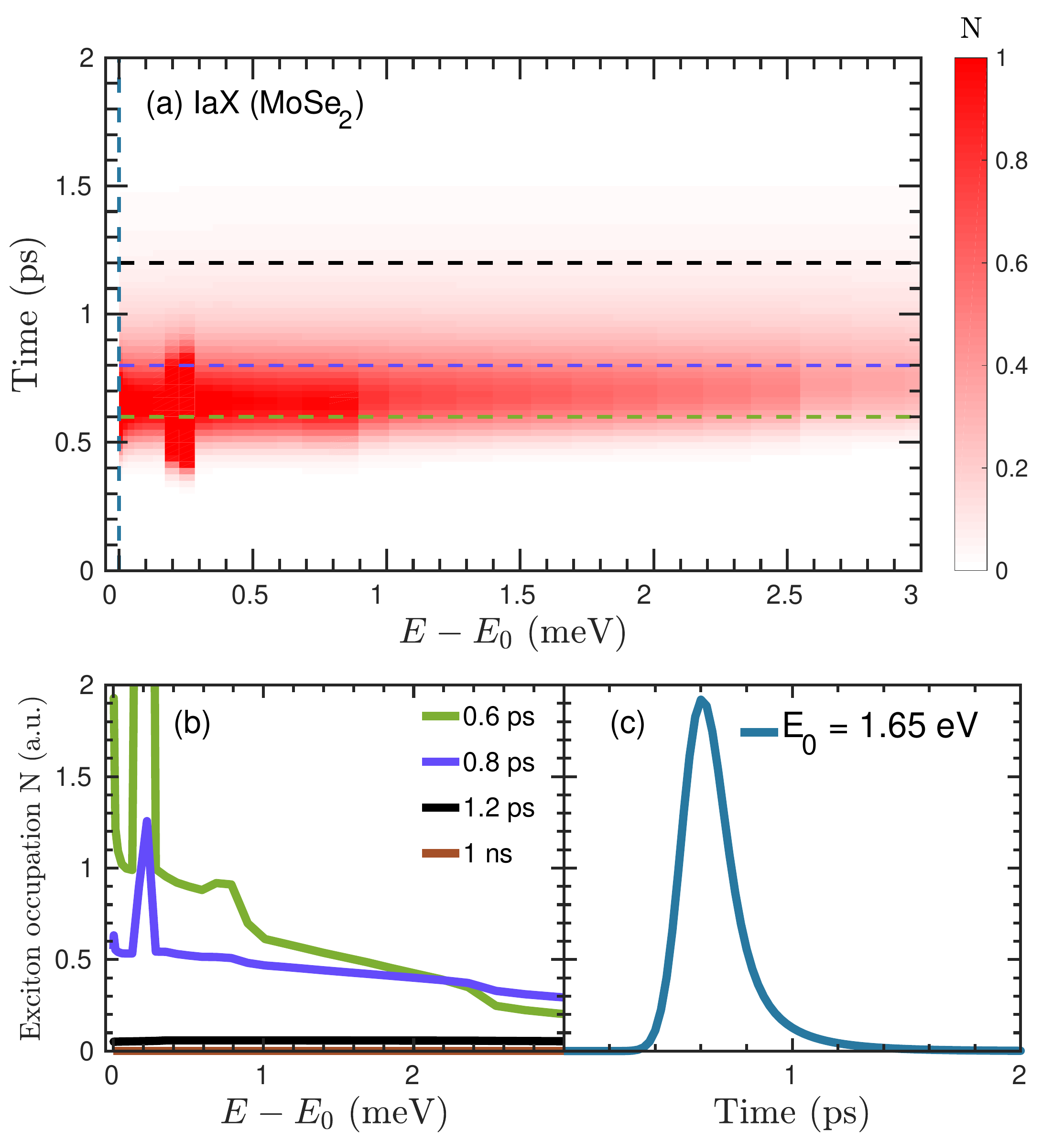}
    \caption{\textbf{Energy- and time-resolved intralayer exciton dynamics.} Temporal evolution of the intralayer exciton (IaX) occupation $N$ at $\unit[77]{K}$ after resonant excitation of the MoSe$_2$ monolayer. (a) The surface plot shows $N$ as a function of time and energy, while (b) and (c) illustrate exemplary snapshots along energy- and time axes (corresponding to dashed lines in (a)), respectively. Through exciton-phonon scattering the system equilibrates into a  Boltzmann-like distribution, which decreases in amplitude due to radiative recombination and interlayer tunneling.}
    \label{fig:NtqIaX}
\end{figure}

Solving the luminescence Bloch equations described above, we can resolve the dynamics of intra- and interlayer excitons. Figure \ref{fig:NtP} shows the temporal evolution of  exciton densities, i.e. momentum-integrated exciton occupations $n=\sum_{\bf Q} N_{\bf Q}$, at the exemplary temperature of $\unit[77]{K}$ (cf. supplementary material for 4$\unit{K}$ and room temperature). The system is excited by a $\unit[100]{fs}$ long Gaussian pulse centered at $\unit[0.5]{ps}$ and a frequency resonant to the intralayer 1s exciton of the  MoSe$_2$ layer. We find that the optically excited coherent excitons  (dashed red line in Fig. \ref{fig:NtP}) decay on a timescale of hundreds of femtoseconds due to radiative emission and exciton-phonon scattering. The latter leads to the formation of incoherent intralayer excitons (IaX) through the so-called polarization-to-population transfer \cite{KKQE, KKSemi, MalteDE, SamuelCascade}. After about $\unit[2]{ps}$, these  intralayer excitons are completely  transferred to interlayer excitons (IeX). 
This occurs through tunneling between energetically resonant states of the two layers and results in a transfer of holes to the WSe$_2$ layer (Fig. \ref{fig:HeteroConcept}). The subsequent phonon-induced relaxation of holes in the WSe$_2$ layer to the valence band maximum effectively traps the holes within that layer, since tunneling back to MoSe$_2$ is energetically forbidden.
The resulting interlayer excitons have lifetimes orders of magnitudes longer than the intralayer excitons, since recombination mechanisms are suppressed due to the spatial separation of Coulomb-bound electrons and holes. 
We predict an interlayer exciton lifetime in the range of hundreds of microseconds at 77K. Experimentally measured sub-nanosecond values \cite{InterKorn} suggest that the investigated radiative decay is not the dominant channel, but rather non-radiative decay e.g. induced by disorder might play the crucial role.
\begin{figure}[t!]
\includegraphics[width=0.85\linewidth]{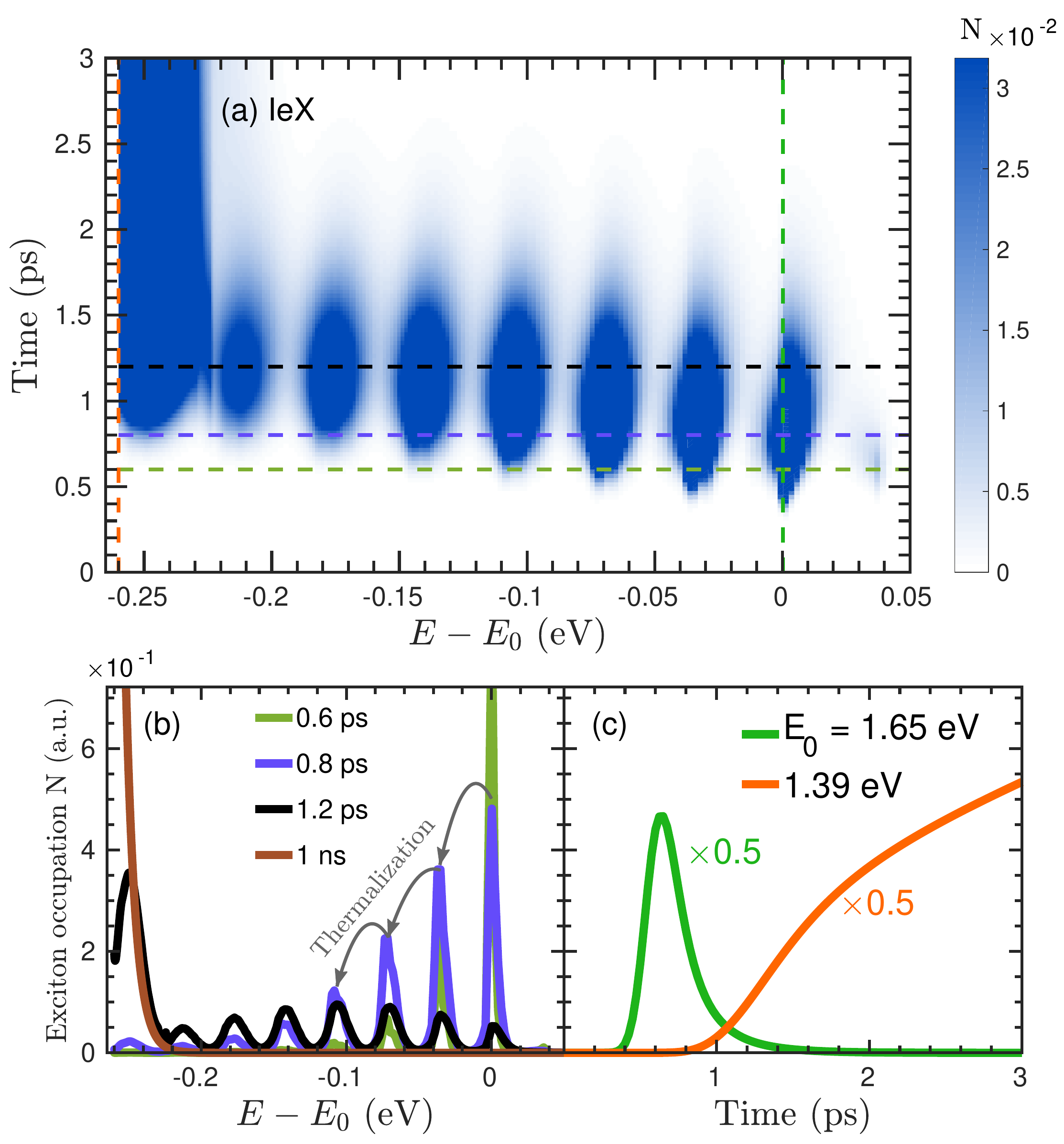}
    \caption{\textbf{Energy- and time-resolved interlayer exciton dynamics.} Same plot as in Fig. \ref{fig:NtqIaX} but showing the time- and energy-resolved dynamics of interlayer excitons (IeX).}
    \label{fig:NtqIeX}
\end{figure}

To provide details of the exciton formation and thermalization process, we now show the time- and energy-resolved dynamics of incoherent intra- and interlayer excitons. Figure \ref{fig:NtqIaX}(a) illustrates how 1s intralayer excitons (IaX) are created in the MoSe$_2$ layer after optical excitation at $\unit[0.5]{ps}$. The process is followed by a phonon-assisted thermalization and tunneling-driven depletion of these excitons. The main features of these dynamics are illustrated in representative snapshots 
along the energy axis at fixed times (Fig.\ref{fig:NtqIaX}(b)).
At $\unit[0.6]{ps}$, a significant number of excitons is still located in the MoSe$_2$ layer. The distribution is in a strong non-equilibrium due to the efficient polarization-to-population transfer, i.e. the excitonic polarization is converted into incoherent exciton occupations with non-vanishing center-of-mass momentum. This is the origin of the observed peaks (green line in Fig. \ref{fig:NtqIaX}(b)) corresponding to the position of intersections of exciton and phonon dispersion. After the coherence has decayed ($\unit[0.8]{ps}$), the occupation starts to thermalize into a Boltzmann distribution (blue line). The occupation within the light cone centered around the exciton dispersion minimum $E_0$=$\unit[1.65]{eV}$ is lowered as a result of radiative recombination. 
Finally, tunneling of holes to the WSe$_2$ layer causes a considerable decay of the intralayer exciton occupation (black line). 

Figure \ref{fig:NtqIeX}(a) illustrates the corresponding dynamics of interlayer excitons (IeX). They first emerge at $\unit[1.65]{eV}$ corresponding to the energy of intralayer excitons in MoSe$_2$. Then, they scatter down towards lower energies predominantly by emitting optical phonons. We observe sharp occupation peaks in constant intervals stemming from optical phonon energies around $\unit[30]{meV}$, cf. Fig. \ref{fig:NtqIaX}(b). The energy conserving nature of the tunneling interaction forces the entire system into a single Boltzmann-like distribution, spanning over both intra- and interlayer excitonic states.  
Finally, the temporal evolution of exciton occupations at fixed energies corresponding to the 1s resonance of intralayer ($\unit[1.65]{eV}$) and interlayer excitons ($\unit[1.39]{eV}$) is shown in Figs. \ref{fig:NtqIaX}(c) and \ref{fig:NtqIeX}(c). These occupations represent the optically active excitons with vanishing center-of-mass momentum. We see how intralayer excitons are created already during the optical excitation and how they decay on a sub-ps timescale due to tunneling of holes into the neighboring WSe$_2$ layer. During this process the corresponding interlayer exciton occupation starts to increase (cf. Fig. \ref{fig:NtqIeX}(c)). It first increases at the energy of $\unit[1.65]{eV}$ resonant to the intralayer exciton (green curve). Driven by scattering with phonons, excitons accumulate at the minimum of the dispersion at $\unit[1.39]{eV}$ (orange curve).

\begin{figure}[!t]
\includegraphics[width=0.85\linewidth]{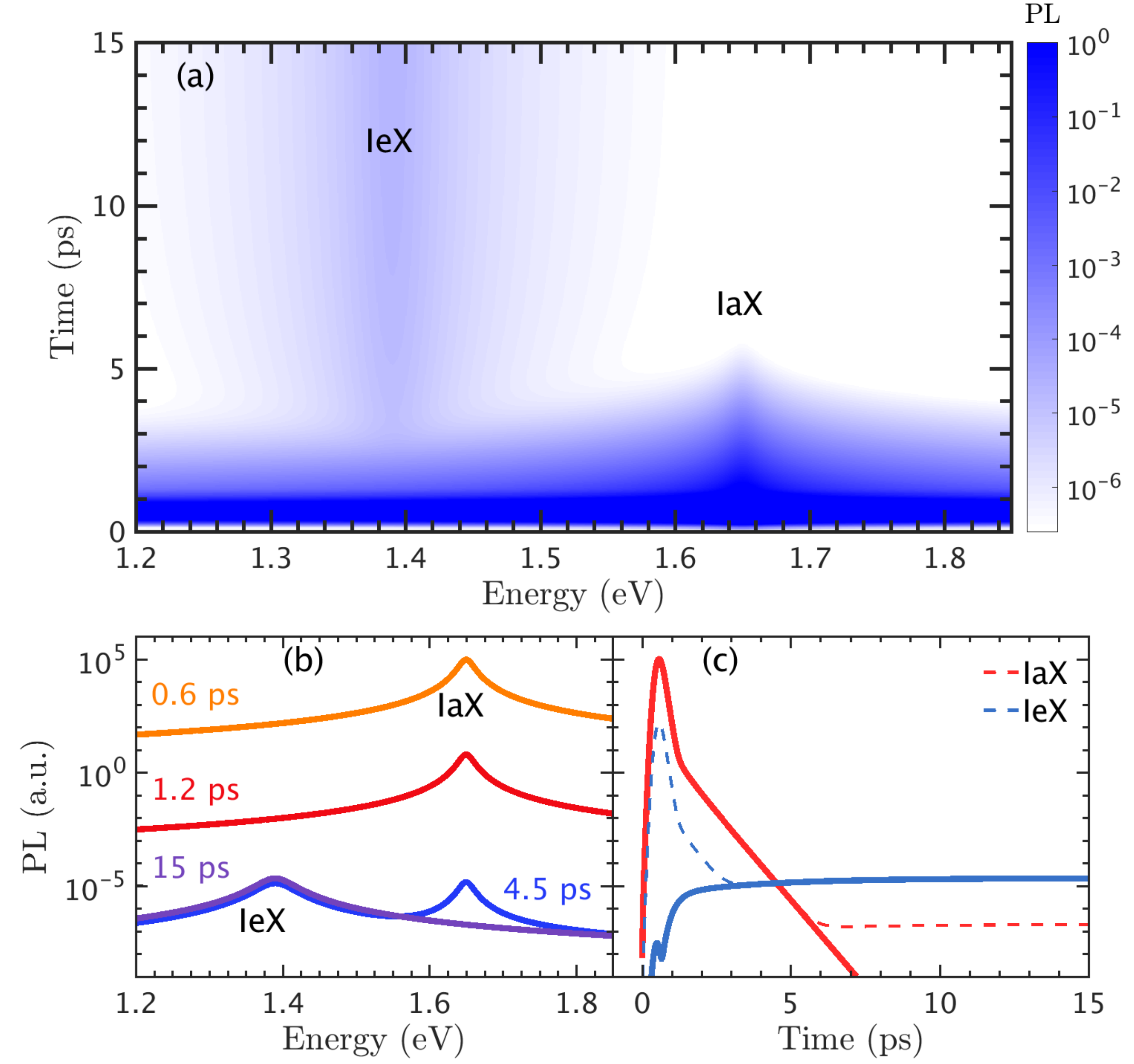}
    \caption{\textbf{Time- and energy-resolved photoluminescence.} PL of the investigated heterostructure at $\unit[77]{K}$ plotted (a) over time and energy (logarithmic) and showing snapshots taken at specific (b) times  and (c) energies. The chosen energies correspond to the position of the interlayer (IeX) and MoSe$_2$ intralayer (IaX) exciton resonance, respectively. The solid lines in (c) represent the signals stemming from the inter- or intralayer emission only, while the dashed lines also include the spectral overlap with the respective other peak.}
    \label{fig:PL}
\end{figure}

The exciton dynamics discussed above determines the light emission from the heterostructure. Figure \ref{fig:PL}(a) shows the time- and energy-resolved PL spectrum at the exemplary temperature of $\unit[77]{K}$ in  (cf. supplementary material for 4 $\unit{K}$ and 300 $\unit{K}$). We find that in the first few ps, the emission from intralayer excitons (IaX) at $\unit[1.65]{eV}$ clearly dominates the PL. However, after approximately $\unit[4.5]{ps}$, the contribution of  interlayer excitons becomes pronounced. To better understand the underlying processes, we show again snapshots at fixed times and energies in Figs. \ref{fig:PL}(b) and (c), respectively. 
The emission stemming from the intralayer exciton at $\unit[1.65]{eV}$ (red line in Fig. \ref{fig:PL}(c)) shows a  maximum PL intensity in the first hundreds of femtoseconds originating from the efficient coherent emission (radiative decay of polarization on a fs timescale). The following slower decay on a time scale of a few ps reflects the decrease of intralayer excitons due to hole tunneling to the WSe$_2$ layer. In this time, interlayer excitons are formed (solid blue line in Fig. \ref{fig:PL}(c)).  After approximately $\unit[4.5]{ps}$, the contribution of the interlayer exciton surpasses the emission from the intralayer exciton.  
Snapshots of the energy-dependent PL along these characteristic times further demonstrate by far most pronounced coherent IaX emission (orange line, $\unit[0.6]{ps}$), the reduced emission due to incoherent IaX (red, $\unit[1.2]{ps}$), equal IaX and IeX emission (blue, $\unit[4.5]{ps}$), and finally the dominant IeX emission (purple, $\unit[15]{ps}$).

Note that after $\unit[15]{ps}$ an equilibrium situation is reached that is characterized by a constant IaX-IeX intensity ratio of approximately 100 (dashed lines in Fig. \ref{fig:PL}(c)).
This is determined  by the ratio of the square of the corresponding optical matrix elements $|M_{\text{IeX}}|^2/|M_{\text{IaX}}|^2$ and exciton occupations $N^{\text{IeX}}_0/N^{\text{IaX}}_0$ within the light cone.
Since the occupations are described by a Boltzmann distribution in equilibrium, we can explicitly calculate the temperature, at which the PL ratio is 1. 
We find that the interlayer exciton emission dominates until approximately $\unit[200]{K}$. Above this temperature, the thermal occupation of energetically higher intralayer excitons increases and considering the significantly larger optical matrix element for intralayer emission, the intralayer peak exceeds the emission from the interlayer exciton. 
Note that the occupation of dark inter-valley excitons will also become important at increased temperatures\cite{MalteDE, GunnarLandscape}, which will also have an influences on the PL intensity. These states are beyond the scope of the current work and will be addressed in a future study.

In conclusion, we have presented a microscopic view on  the dynamics of inter- and intralayer excitons in van der Waals heterostructures.  Solving the luminescence Bloch equations, we reveal the time- and energy-resolved processes behind the formation, thermalization, and decay of interlayer excitons. We predict that tunneling of holes from the optically excited  into the neighboring  layer is the dominant formation channel occurring on a sub-picosecond timescale. Although the radiative recombination is strongly quenched due to the spatial separation of charge carriers, we show that at temperatures below $\unit[200]{K}$ the photoluminescence is dominated by interlayer excitons. 
The gained insights will trigger new experimental studies on van der Waals heterostructures. In particular, the predicted formation dynamics of interlayer  excitons can be experimentally addressed by pumping and probing the intralayer exciton transition in different layers, where a clear bleaching is expected due to the efficient interlayer tunneling. 

This project has received funding from the European Union's Horizon 2020 research and innovation programme under grant agreement No 696656. Furthermore, we acknowledge financial support from the Swedish Research Council (VR), the Stiftelsen Olle Engkvist, and the Deutsche Forschungsgemeinschaft (DFG) through SFB 951 and the School of Nanophotonics (SFB 787). 

\bibliographystyle{apsrev4-1}
\bibliography{references.bib}

\end{document}